\documentclass[useAMS,usenatbib]{mn2e}

\newcommand{\uk}{{\rm $\mu$K\,} }
\newcommand{\dg}{$^\circ$\,}

\title[]{The Geometric Calibration of the {\it Planck} satellite using point-source observations}
\author[D. L. Harrison]{D. L. Harrison$^{1}$\thanks{E-mail: dlh@ast.cam.ac.uk (DLH)} and F. van Leeuwen$^{1}$ \\
$^{1}$ Institute of Astronomy, Madingley Road, Cambridge, CB3 0HA, UK\\}
\begin{document}

\date{}

\pagerange{\pageref{firstpage}--\pageref{lastpage}} \pubyear{2005}

\maketitle

\label{firstpage}

\begin{abstract}

The geometric calibration of the {\it Planck} satellite is investigated, specifically those parameters which require the use of the science data for their extraction. Methods for the recovery of these geometric-calibration parameters from point source detections in the science data are presented, together with the accuracies which may be achieved. These methods apply to the a posteriori evaluation of these parameters using all the mission data, and may be incorporated into the initial stages of the construction of the {\it Planck} final compact source catalogue. It is found that this process achieves the pointing requirements, provided that the geometric-calibration parameters vary only slowly in time. Indeed the errors in the pointing reconstruction due to the geometric calibration parameters may be seen to approach those due to the star tracker.

\end{abstract}


\section{Introduction}
\label{intro}

{\it Planck} is a European Space Agency satellite designed to produce high-resolution temperature and polarisation maps of the CMB. It is equipped with detectors sensitive to a wide range of frequencies from 30 to 857 GHz, split between two instruments, HFI and LFI, the high and low frequency instruments, respectively. The full-sky coverage and frequency range provided by {\it Planck} will also enable the construction of a unique compact source catalogue. Below 100~GHz it will be a significant improvement over that of WMAP, (\cite{bennett03}), in terms of sensitivity and hence in the number of sources detected, while above 100~GHz it will be the only full-sky survey for many years to come.

This paper investigates the possiblity of recovering the geometric-calibration parameters, defined in section~\ref{geoCal_param}, as part of the initial stages of the construction of this source catalogue. Previous work, (\cite{harrison04}), on the reconstruction of the geometric-calibration parameters concentrated on the recovery of the boresight and roll angle parameters during the course of the mission; the recovery of the reference phase was not discussed, nor has it been discussed elsewhere. While the monitoring and evaluation of these parameters is crucial over the lifetime of the mission, it is expected that much higher accuracies are achievable with an a posteriori evaluation of these parameters using the entirety of the mission data. Hence it is expected that the construction of the final compact source catalogue, FCSC, will provide the definitive values of the geometric-calibration parameters.

An overview  of the FCSC is presented in Section~\ref{overview_fcsc}. The geometric-calibration parameters are defined in Section~\ref{geoCal}, together with a discussion on their use in the pointing reconstruction of the {\it Planck} field-of-view and the accuracy requirements of the pointing reconstruction are assessed. The simulations generated to assess the performance of the geometric-calibration methods, outlined in Section~\ref{methods_section}, are discussed in Section~\ref{simul_section}. The results of this analysis are presented in Section~\ref{results_section}, where we show that it is possible to recover these parameters to the required accuracies using solely the bright extragalactic point sources, provided there is only a slow linear variation in the value of these parameters over the course of the mission.

\section{An Overview of the Final Compact Source Catalogue}
\label{overview_fcsc}

The Final Compact Source Catalogue, FCSC, should not be confused with the Early Release Compact Source Catalogue, ERCSC. The ERCSC will be released approximately 1 year after the start of the routine observations. Its primary purpose being to allow rapid follow-up observations from Herschel and ground-based millimetre instruments. This catalogue will be produced using the frequency-channel maps from the data corresponding to the first full-sky coverage, this is expected to be the first $\sim$ 7 months worth of data. The requirement on the release date, however, curtails the focus of the ERCSC to the brightest sources. Further details of the ERCSC implementation plan may be found in \cite{ganga02}. The FCSC will comprise of the full mission data for {\it Planck} and be a significant improvement over the ERCSC.

 The primary goal in the construction of the FCSC is to ensure the internal consistency and accuracy of the catalogue and final frequency maps. This goal has motivated the division of the catalogue construction into four major stages.

\begin{enumerate}
\item{Stage 1a: The accumulation of the detections from bright point sources in the ring data, where the ring data corresponds to a single spin axis positioning, is used to find positions for these point sources and any deviations from the current bestfit values for the geometric-calibration parameters. It is this stage which forms the main focus of this paper.}
\item{Stage 1b: Solar system objects with their highly accurate positional data also allow the recovery of the geometric-calibration parameters. Given the relative motion of these bodies and the {\it Planck} satellite the methods employed are some what different from those in Stage~1a. Hence the discussion of the methods used in Stage~1b is reserved for paper~II, {\it The Geometric Calibration of the Planck satellite using Solar System Objects} }
\item{Stage 2 \& 3: These stages involve the treatment of the intermediate-source detections on rings and the source detections in maps, respectively. It is envisaged that for the source detections from maps, it may be necessary to return to the ring data to ensure that maximum accuracy and consistency will be attained. The discussion of the methods involved in these stages of construction is left to paper~III, on the final compact source catalogue}
\end{enumerate}

\section{Geometric Calibration}
\label{geoCal}

{\it Planck} will be inserted into a Lissajous orbit around the second Lagrange point of the Earth-Sun system, spinning about its axis once per minute. The line of sight being almost perpendicular to the spin axis, hence the detectors will almost follow a great circle. The spin axis will nominally be repositioned every hour, and the roughly 60 or so circles corresponding to a single spin axis positioning may be binned together to form a ring. The nominal spin axis passes through the centre of the solar panels and is directed away from the sun, thus maintaining the rest of the satellite in a cone of shadow produced by the solar panels.  The scanning strategy is determined by the sequence of the nominal positions of the spin axis over the course of the mission. 

The geometric calibration is the process whereby all the line-of-sight positions of the detectors are recovered, at any time during the observations. The relationship between the pointing and time may depend on multiple parameters, discussed in full in \cite{leeuwen02}; the star tracker will provide many of these parameters. However, as will be discussed in Section~\ref{pointingRecon}, there is an uncertainty between the line-of-sight of the star tracker and that of the focal plane, this will produce offsets in the values of a few of the geometric-calibration parameters. These offsets may only be recovered with the science data, and it is those geometric-calibration parameters which require this calibration with the science data that we discuss here.

\subsection{The Geometric-Calibration Parameters}
\label{geoCal_param}

\begin{figure}
\begin{center}
\setlength{\unitlength}{1cm}
\begin{picture}(10,8)(0,0)
\put(-1.5,10){\includegraphics{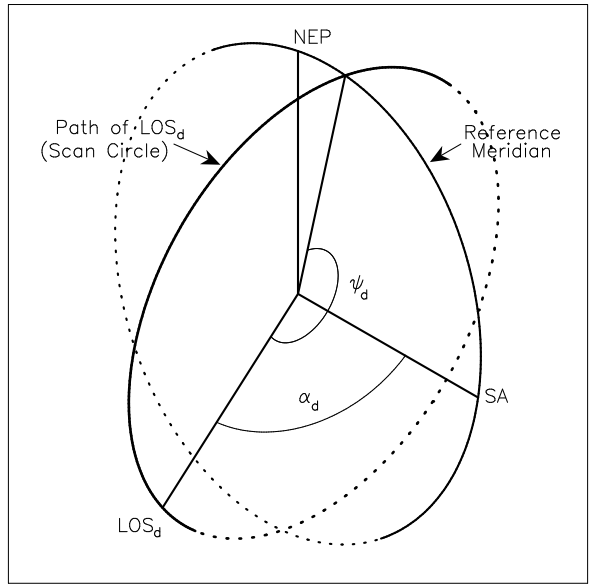}}
\end{picture}
\end{center}
\caption[]{The figure shows the two angles which describe the position of the line-of-sight of a detector, ${\rm LOS_d}$, with respect to the spin axis position, SA, and the North Ecliptic Pole, NEP. The angle between the spin axis and the ${\rm LOS_d}$ is given by the opening angle to the detector, $\alpha_d$. The opening angle to the detector from the spin axis position describes the path followed along the sky by the ${\rm LOS_d}$. The position of the ${\rm LOS_d}$ along this path is given by the phase, $\psi_d$. The zero point for the phase may be determined by the intersection point of the great circle which connects the spin axis position and the NEP and the path of the scan circle. This great circle may be defined as the reference meridian.}
\label{los_fig}
\end{figure}
The position of the line-of-sight of a detector with respect to the spin axis position may be described in terms of two angles, which are shown in Figure~\ref{los_fig}. The first, the opening angle, is the angle between the spin axis position and the line-of-sight of the detector. The opening angle defines the path described by the detector, the scan circle. The second angle, the phase, defines the position along the scan circle from a given reference point. This reference point may be given by the intersection point of the scan circle and the great circle connecting the spin axis position with the North Ecliptic Pole, NEP. This great circle, shown in Figure~\ref{los_fig}, will be referred to as the reference meridian. Each repositioning of the spin axis defines its own reference meridian.
\begin{figure}
\begin{center}
\setlength{\unitlength}{1cm}
\begin{picture}(10,8)(0,0)
\put(-1.5,10){\includegraphics{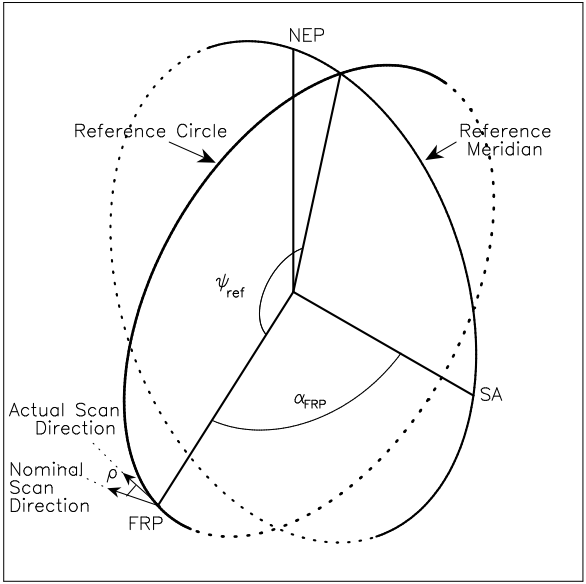}}
\end{picture}
\end{center}
\caption[]{The figure shows the geometric-calibration parameters which allow a description of the position of the field of view, FOV, with respect to the spin axis position, SA. The boresight angle, $\alpha_{FRP}$ is the angle between the FRP and the spin axis. The rotation of the focal plane around the FRP with respect to the nominal scan direction is given by the roll angle, $\rho$. The reference phase, $\psi_{ref}$, is the value of the initial phase at the point at which the FRP crosses the reference point, as defined by the intersection point of the reference circle and reference meridian closest to the NEP. The initial phase is measured from the first point observed on the reference circle.}
\label{geo_fig}
\end{figure}
\begin{figure}
\begin{center}
\setlength{\unitlength}{1cm}
\begin{picture}(8,8)(0,0)
\put(8.85,0){\includegraphics{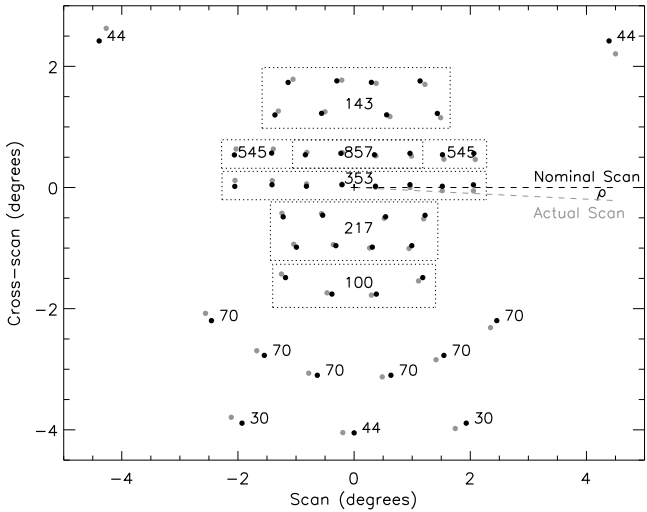}}
\end{picture}
\end{center}
\caption[]{The figure shows the positions of the line-of-sight of the {\it Planck} detectors, filled circles, relative to the FRP which is shown by the cross. The black circles are the positions for the nominal scan direction, whereas the grey circles illustrate the effects of a non-zero roll angle on the detector positions. The dotted lines enclose detectors belonging to the same frequency channel.}
\label{roll_fig}
\end{figure}

If we assume the focal-plane geometry, as defined by the relative locations of the detectors to the position of a fiducial reference point, FRP, defined somewhere central in the focal plane, remains fixed, then instead of requiring two angles per detector we may describe the positions of all the detectors in terms of just three angles.  These three angles, shown in Figure~\ref{geo_fig}, are the opening angle to the FRP, otherwise known as the boresight angle, $\alpha_{FRP}$, the phase $\psi_{FRP}$ of the FRP, determined from the reference phase, $\psi_{ref}$, as measured from the reference point defined by the reference meridian, and the roll angle, $\rho$, which is given by the rotation of the focal plane around the FRP, relative to a nominal scan direction defined within the focal-plane geometry, as illustrated in Figure~\ref{roll_fig}.

The position of the line-of-sight of a detector may now be described in terms of the nominal values for these angles, the phase of the $d^{th}$ detector is given by:
\begin{equation}
\label{def_detPhase_eqn}
\psi_d=\psi_{FRP}+x_d(\rho)
\end{equation} and similarly the opening angle to the detector by:
\begin{equation}
\label{openAngle_eqn}
\alpha_d=\alpha_{FRP}+y_d(\rho)
\end{equation}  where $x_d$ and $y_d$ are respectively, the scan and cross-scan positions of the $d^{th}$ detector with respect to the FRP and may be determined by the focal-plane layout and the roll angle. The scan and cross-scan positions of a detector for a given roll angle, $\rho$, may be found using the position given for the detector in the focal plane, $(x_{d,0}, y_{d,0})$, which corresponds to the scan and cross-scan positions of the detector in the case of zero roll angle:
\begin{equation}
\label{rotation_eqn}
\left(\matrix{x_d \cr y_d}\right) = \left(\matrix{ \cos \rho  & \sin \rho \cr -\sin \rho  & \cos \rho }\right) \left(\matrix{x_{d,0} \cr y_{d,0}}\right)
\end{equation}

The actual values of the geometric-calibration parameters, however, may deviate from their nominal values, as will be discussed in Section~\ref{pointingRecon}. It is therefore useful to express the offset in the phase, $\Delta \psi_d$, and opening angle, $\Delta \alpha_d$, to a detector in terms the differences between the nominal and actual geometric-calibration parameters using equations~\ref{def_detPhase_eqn} to~\ref{rotation_eqn}:

\begin{eqnarray}
\label{offset_dep_eqn}
\Delta \psi_d &=& \delta \psi_{FRP} + x_{d}(\rho)\left(\cos(\delta \rho)-1\right) + y_{d}(\rho)\,\sin(\delta \rho), \nonumber\\
\Delta \alpha_d &=& \delta \alpha_{FRP} - x_{d}(\rho)\,\sin(\delta \rho) + y_{d}(\rho)\left( \cos(\delta \rho)-1 \right).
\end{eqnarray}

The dependence of the recovered position of the line-of-sight of a detector may now clearly be seen in equation~\ref{offset_dep_eqn}; the position along the scan is affected by the reference phase and the roll angle, whereas the position in the cross-scan direction is affected by boresight and roll angles. Information on the scan-phase correction $\Delta \psi_d$ may be obtained from the measurements of $\psi_d$ for point-source transits. The correction to the opening angle $\Delta \alpha_d$ may be derived from the distribution of intensities of the point-source transits. These corrections may be used to evaluate the offsets in the geometric-calibration parameters, this analysis is the subject of Section~\ref{methods_section}.

\subsection{Pointing Reconstruction}
\label{pointingRecon}

\begin{figure}
\begin{center}
\setlength{\unitlength}{1cm}
\begin{picture}(10,8)(0,0)
\put(-0.25,0){\includegraphics{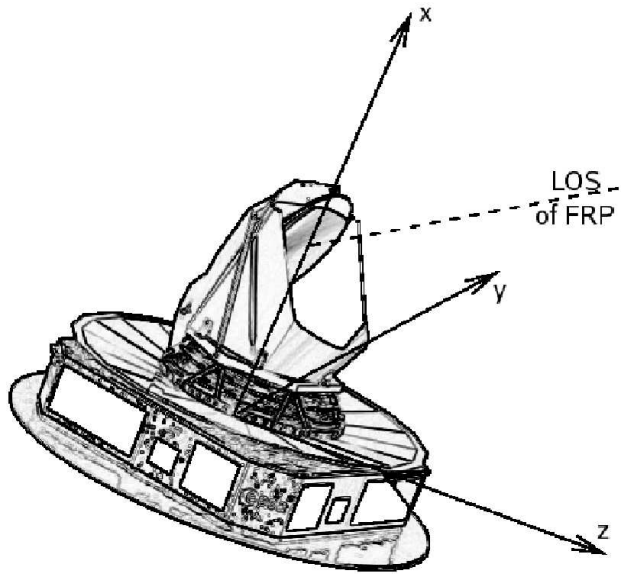}}
\end{picture}
\end{center}
\caption[]{The figure shows the Satellite Reference System, SRS, which is defined in terms of the principle axes of the {\it Planck} satellite. The $x$-axis is aligned with the nominal spin axis, the line-of-sight of the FRP lies in the $xz$-plane, and the $y$-axis completes the right-handed triad.}
\label{planck_SRS_fig}
\end{figure}

The pointing reconstruction of the satellite is achieved by the recovery of the the relationship between the line-of-sight of the star tracker and that of the focal plane. 

It is helpful to define two reference frames for {\it Planck}, the Satellite Reference System (SRS) and the Inertial Reference System (IRS). The origin of both systems is the centre of gravity, CoG, of the satellite. The SRS is aligned with the principle axes of the satellite, and is shown in Figure~\ref{planck_SRS_fig}. The $x$-axis is defined by the nominal spin axis which passes through the centre of the solar panels, and the CoG, and is directed away from the Sun. The $z$-axis is orthogonal to the $x$-axis and lies in the plane defined by the $x$-axis and the projected line-of-sight of the FRP. The $y$-axis then completes this right-handed triad. It is necessary to define a second reference system, the Inertial Reference System, IRS, in which the inertia tensor is diagonal. The $x$-axis is now the actual spin axis, which is defined by the largest principle axis as determined by the principle moments of inertia of the satellite.

These two reference systems are connected by a time varing rotation matrix which must be recovered in flight. As the mission progresses and consumables are depleted this affects the inertia tensor of the satellite and hence the axis of rotation, changing the relationship between the SRS and the IRS. The star tracker should recover this time varing relationship. However, the exact relationship between the star tracker reference system and the SRS is not known. This lack of an exact relation between the star tracker and the SRS and IRS will produce uncertainties in the reconstruction of the actual line-of-sights for the detectors. These uncertainties have been estimated to be $\sim$~1.3 arc-minutes, \cite{chlewicki04}, and may be described in terms of offsets in the geometric-calibration parameters, which must be recovered from the science data in order to meet the accuracy requirements discussed in Section~\ref{acc_reqments}.

The relationships between the focal plane and the star tracker to the SRS, although nominally constant may indeed have a slow variation with time due to thermal effects. The offsets in the geometric-calibration parameters which must be established from the science data, may therefore be time dependent. Additional time variation may arise if the star tracker fails to completely establish the time dependence between the SRS and the IRS. The expected magnitude of these effects are unknown, but are thought to be less than 1 arcmin.

A more detailed discussion of the reference systems and the attitude analysis may be found in \cite{leeuwen02}.

\subsection{Pointing Accuracy Requirements}
\label{acc_reqments}

\begin{figure}
\begin{center}
\setlength{\unitlength}{1cm}
\begin{picture}(7,7)(0,0)
\put(8.5,0){\includegraphics{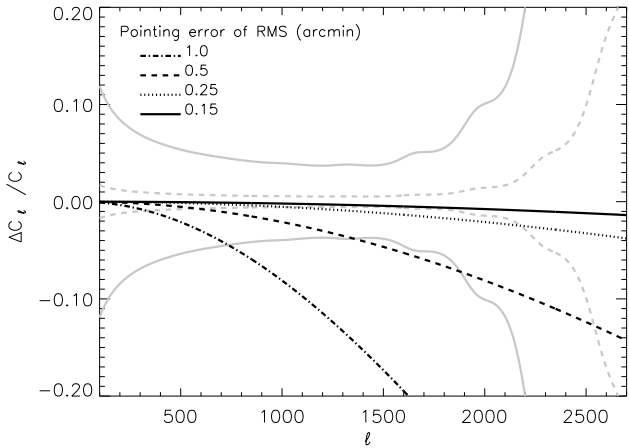}}
\end{picture}
\end{center}
\caption[]{The figure shows the unsubtractable errors on the $C_{\ell}$s which are dominated by cosmic variance to the left and the finite beam size and noise to the right. The solid grey curves enclose the region corresponding to the error on each individual multipole, whereas the dashed grey curves correspond to the error on multipole bins of width 50. The black curves correspond to the additional errors on the reconstruction of the $C_{\ell}$s due to unknown random pointing errors. A beam of FWHM 5\arcmin, with a noise of 10.3\uk per beam and a fractional sky coverage of 0.7 have been assumed.}
\label{acc_fig}
\end{figure}

Figure~\ref{acc_fig} was created in order to assess the accuracy to which the pointing must be recovered to avoid compromising the reconstruction of the $C_\ell$ power spectrum, which describes the magnitude of the fluctions in the CMB as a function of angular scale and is the primary science goal of the {\it Planck} satellite. It shows the effect of unknown random pointing reconstruction errors on the recovered $C_{\ell}$ values. The solid grey curves correspond to the error on individual multipoles, whereas the dashed grey curves correspond to the error on multipole bins of 50. The shape of the grey curve is determined at low multipoles, on the left of the figure, by the sample and cosmic variance. On the right at high multipoles, however, the error is due to noise and the finite size of the beam. The decrease in the sensitivity of the beam to the higher multipoles may be accounted for at the cost of the exponential increase in the noise at the high $\ell$ values; as the noise is undiminished by the beam size. The series of black curves in Figure~\ref{acc_fig} show the effect of unknown random pointing reconstruction errors. The pointing uncertainties result in an effective smearing of the beam, resulting in a larger effective beam. This produces the reduction in the sensitivity to the higher multipoles, as seen in Figure~\ref{acc_fig}. A more detailed discussion of how this figure is obtained, may be found in \cite{harrison04}. Ideally, the errors in the reconstruction of the pointing should be such that the additional errors introduced are less than that of the unsubtractable noise. This places an upper limit on the total pointing error of 9~arcseconds. 

It should be noted that correlated errors in the pointing will not produce effects as large as those for pointing errors of the same magnitude which are random. The effect of smearing the beam predominately in one direction is to increase the sensitivity to higher $\ell$ multipoles as compared to the uniform smearing shown in Figure~\ref{acc_fig}. The pointing accuracy requirement generated by assuming unknown random pointing errors should therefore exceed any requirements found from assuming correlated errors, and will represent the tightest constraints on the pointing accuracy required.  

\section{Methods for the geometric calibration}
\label{methods_section}

As discussed in Section~\ref{pointingRecon} there will be an offset between the line-of-sight, LOS, as determined by the Star Tracker and the actual LOS. It is this offset which produces the systematic offsets between the actual and nominal values of the reference phase, boresight and roll angles, which may only be recovered using the science data.

The methods for evaluating the geometric-calibration parameters therefore, need only assess the offset in the value of the parameters from their nominal values. These methods should also be able to cope with a slow variation in the values of these offsets over the course of the mission, due to thermal effects, as discussed in Section~\ref{pointingRecon}. These offsets must be recovered to accuracies which will allow the total pointing error to remain below 9 arcseconds, as discussed in Section~\ref{acc_reqments}.

As discussed in Section~\ref{geoCal_param}, offsets in the reference phase and the boresight angle produce errors in the recovered position which are orthogonal, and hence may be solved for independently. The offset in the roll angle, however,  produces errors in position, in both the scan and cross-scan directions. In practice, however, the roll angle may be solved together with the reference phase independently from the boresight angle, as will be discussed in Section~\ref{methods_refPhase}. The calibration of the opening angles to the detectors and hence the boresight angle is discussed in Section~\ref{methods_openAng}.

\subsection{Reference Phase and Roll Angle}
\label{methods_refPhase}

\begin{figure}
\begin{center}
\setlength{\unitlength}{1cm}
\begin{picture}(8.5,8.5)(0,0)
\put(0,0){\includegraphics{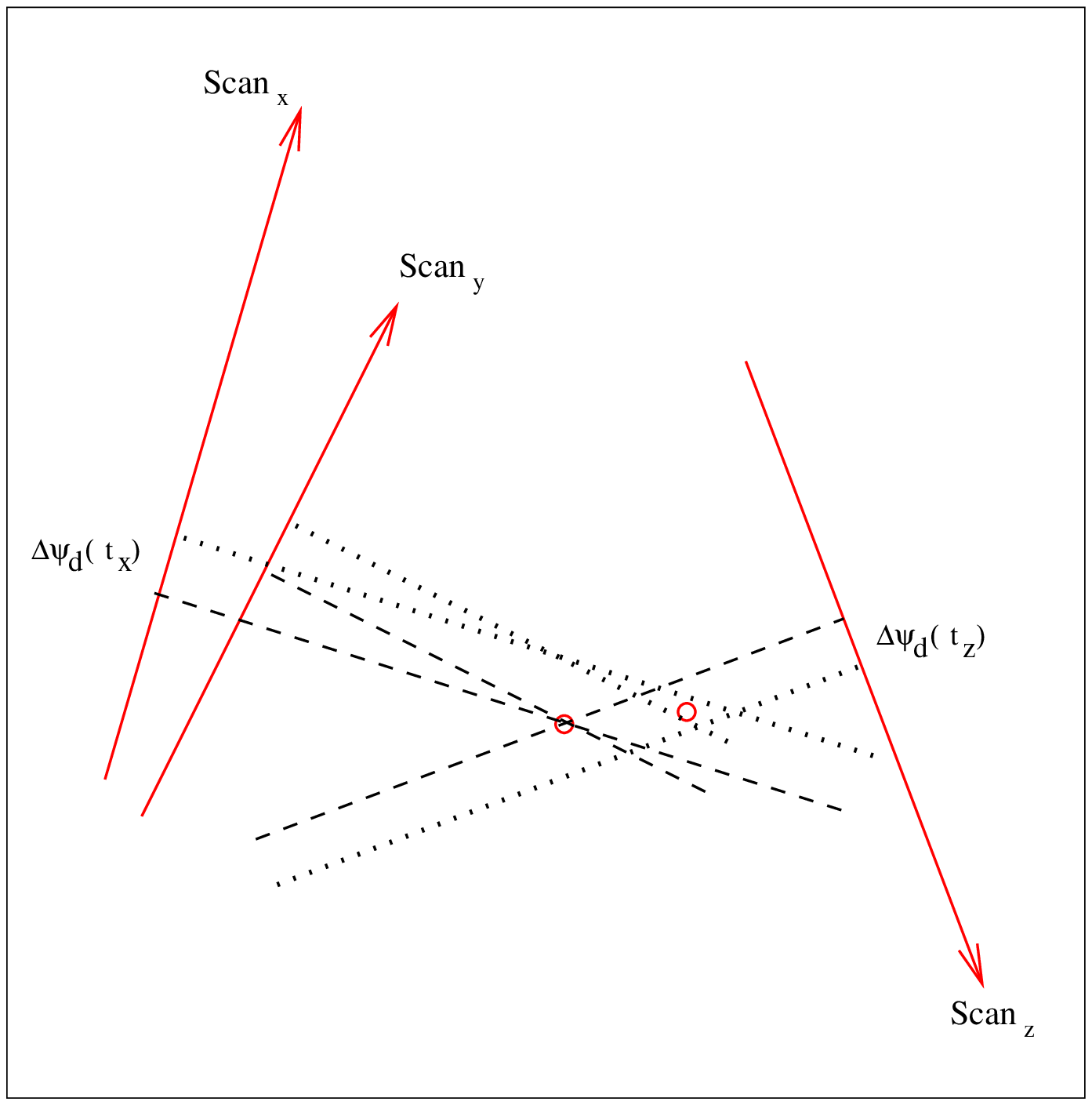}}
\end{picture}
\end{center}
\caption[]{This figure shows the evaluation of the position of a point source, with, dashed lines, and without, dotted lines, including corrections to offsets in the reference phase or roll angle. The minimisation of the residuals between the expected phase of the source given its position and the observed phase of the detection allows an assessment of the offsets in the reference phase and roll angle as a function of time.}
\label{intersect_fig}
\end{figure}

The position of a point source in the scan direction may be constrained by the phase of the detection. Detections on multiple non-parallel scans which correspond to the same point source, may therefore be used to constrain the position of the point source on the sky. This process is illustrated in Figure~\ref{intersect_fig}, where the solid arrows represent the scan directions. The position of the point source may be found from the intersection point of lines orthogonal to the scan direction extending from the phase of each detection. Figure~\ref{intersect_fig} shows these lines for the cases of no offsets in phase, dashed lines, and unreconstructed time dependent offsets, dotted lines. As seen in Figure~\ref{intersect_fig} any unaccounted for offset in phase will affect the recovered position of the point source. Offsets in the opening angles to the detectors, however, will not affect the recovered position. Offsets in the opening angles only affect the angular separation between the recovered position and the apparent position of the scan. The offsets in the phase of the detections, therefore, allow an offset in the roll angle to be discovered independently of an offset in the boresight angle. An offset in the roll angle is equivalent to a different reference phase offset for every detector, hence an offset in the reference phase may be distinguished from an offset in the roll angle. The minimisation of the residuals between the phase observed for the detection and the phase expected for the point source based on its recovered position, therefore provides a mechanism whereby the offset in the reference phase and roll angle may be assessed. The methods presented here are similar to those employed in the sphere reconstruction for the {\it Hipparcos} satellite, \cite{esa97} and~\cite{lindegren92}.

The residual phase, $\Delta \psi_{ij}$, for the $i^{th}$ detection corresponding to the $j^{th}$ point source, may now be defined as the difference between the expected phase for $i^{th}$ detection given the position of the $j^{th}$ point source, $\psi_{ij}$, and the observed phase of the detection, $\psi_i$.  
\begin{equation}
\label{abscissa_eqn}
\Delta \psi_{ij}= \psi_{ij} - \psi_{i}
\end{equation} The expected phase, $\psi_{ij}$, may be evaluated using:
\begin{equation}
\label{refPhase_roll_dep}
\psi_{ij}= \psi_{FRP_{ij}} + x_d(\rho_i)
\end{equation} where $x_d$ is the scan phase of the detector, in which the $i^{th}$ detection occurs, with respect to the FRP and is dependent on the current value of the roll angle at the time of this detection, $\rho_i$. The expected phase of the FRP for the $i^{th}$ detection of the $j^{th}$ point source, $\psi_{FRP_{ij}}$, may be calculated using:
\begin{equation}
\label{phase_pos}
\tan (\psi_{FRP_{ij}})=\frac{\sin (\lambda_{SA_i}-\lambda_j)}{\cos \beta_{SA_i} \tan \beta_j -\sin \beta_{SA_i} \cos(\lambda_{SA_i}- \lambda_j)}
\end{equation} where $(\lambda_{SA_i}, \beta_{SA_i})$ is the spin axis position corresponding to the $i^{th}$ detection and  $(\lambda_j,\beta_j)$ is the position of the $j^{th}$ point source. An initial position for the $j^{th}$ point source, $(\lambda_j,\beta_j)$, may be found by minimimizing the residual phases for all the detections of the $j^{th}$ point source, hence solving:
\begin{equation}
\label{initial_pos_eqn}
\frac{\partial \sum_i (\delta \psi_i^2)}{\partial \lambda_{j}}=0\,\,{\rm and}\,\,
\frac{\partial \sum_i (\delta \psi_i^2)}{\partial \beta_{j}}=0
\end{equation} where the summation is over all the detections of the $j^{th}$ point source and  $\delta \psi_i$ is the residual in the phase of the $i^{th}$ detection which may be found using the relationship between small changes in the phase of a detection and the resultant change in position, $(\delta \lambda_i, \delta \beta_i)$ :
\begin{eqnarray}
\label{del_phase_del_pos}
\delta \psi_i  & = & \left( \frac{\sin \beta_{SA_i} -\cos \alpha_i \sin \beta_i}{\sin^2 \alpha_i} \right) \delta \lambda_i \nonumber \\
& & \mbox{} -\left( \frac{\cos \beta_{SA_i} \sin(\lambda_{SA_i}-\lambda_i)}{\sin^2 \alpha_i} \right) \delta \beta_i
\end{eqnarray} where $\alpha_i$ is the current value, at the time of the $i^{th}$ detection, of the opening angle to the detector in which the $i^{th}$ detection occurs, and  $\delta \lambda_i = \lambda_{j}-\lambda_i $ and $\delta \beta_i = \beta_{j}-\beta_i $, where $(\lambda_i,\beta_i)$ is the position of the detection.

Once an initial position for the $j^{th}$ point source has been found, the expected phase of the detection, $\psi_{ij}$, and hence the residual phase, $\Delta \psi_{ij}$, may be evaluated. The residual phase, however, may also be related to changes in the expected phase due to changes in the position of the $j^{th}$ point source, and the residuals in the phase of the detection, $\Delta \psi_d $:
\begin{equation}
\label{refPhase_eqn}
\Delta \psi_{ij}=\delta \psi_{ij} + \Delta \psi_d .
\end{equation} Using equation~\ref{offset_dep_eqn} and under the assumption that the offset in the roll angle is small, $\Delta \psi_d $ may be expressed as:
\begin{equation}
\label{residPhase_eqn}
\Delta \psi_d = -\delta \psi_{ref}(t_i) + y_d (\rho)\, \delta \rho (t_i)
\end{equation} where the offsets in the reference phase and roll angle, may be expressed as a constant offset and a term defining the linear drift in time,
\begin{eqnarray}
\label{dif_refPhase_eqn}
\delta \psi_{ref}(t_i) &= &\left( \psi_{ref_0} +\psi_{ref_1} t_i \right), \nonumber \\
\delta \rho (t_i) &= &\left( \rho_0 +\rho_1 t_i \right).
\end{eqnarray}

Equation~\ref{refPhase_eqn} may now be written in terms of changes to the point source position and the offsets in the reference phase and roll angle, using equations~\ref{del_phase_del_pos} and equation~\ref{residPhase_eqn}.

\begin{eqnarray}
\label{lsq_eqn}
\Delta \psi_{ij}  & = & \Lambda_{ij} \delta \lambda_{j} - B_{ij} \delta \beta_{j} - \left( \psi_{ref_0} +\psi_{ref_1} t_i \right) \nonumber \\
& & + y_d (\rho)\,\left( \rho_0 +\rho_1 t_i \right)\nonumber \\
{\rm where,} & & \nonumber \\
\Lambda_{ij} & = & \left( \frac{\sin \beta_{SA_i} -\cos \alpha_i \sin \beta_{j}}{\sin^2 \alpha_i} \right) \nonumber \\
B_{ij} & = & \left( \frac{\cos \beta_{SA_i} \sin(\lambda_{SA_i}-\lambda_{j})}{\sin^2 \alpha_i} \right) 
\end{eqnarray} A similar equation may be written for every detection, giving $i$ equations and $2 \times j+4$ unknowns. The corrections to the point source positions and the values for the systematic offset and drift in the reference phase and roll angle may then be extracted by a nonlinear least squares analysis.

\subsection{The Boresight Angle}
\label{methods_openAng}

As discussed in Section~\ref{geoCal_param}, the opening angle to a detector is given by the angle between the line-of-sight of the detector and the spin axis position. The opening angles to the detectors are determined by the focal-plane layout and the boresight and roll angles, hence if the offset in the roll angle has been accounted for, and the focal-plane layout is known, any remaining offset in the opening angles will be due to an offset in the boresight angle. 

Once the values of the reference phase, roll angle and the point source positions have been evaluated, as above, the positions of the point sources may be used to evaluate the value of the ordinate for each detection, where the ordinate is angular separation, in the cross-scan direction, between the point source and the path of LOS of the detector.

The detections corresponding to the $k^{th}$ transit of the focal plane, in the cross-scan direction, of the $j^{th}$ point source are accumulated for each of the detectors. It is then possible to find the value of the ordinates which correspond to the peak of the transits, in the cross-scan direction, for each detector in which the $k^{th}$ transit was observed. If there is no offset in the opening angle to the $d^{th}$ detector the value of the ordinate at the peak of the transit will be zero. Hence the offsets in the opening angles may be determined, at the epoch of each transit, by the analysis of the cross-scan transits of the point sources.

The offsets in the opening angle to the detectors for each transit of the focal plane may then be used to find the offset in the boresight angle at the epoch of the transit, $\delta \alpha_{FRP}(t_k)$. As with the reference phase and roll angle parameters, this offset may be expressed as a constant offset and a linear drift in time:
\begin{equation}
\label{bore_drift_eqn}
\delta \alpha_{FRP}(t_k)  =  \delta \alpha_{FRP_0} + \delta \alpha_{FRP_1} t_k\end{equation} where $t_k$ is the epoch of the $k^{th}$ transit.  By assessing the value of $\delta \alpha_{FRP}$ at every available epoch, epochs at which the focal plane transited a sufficiently bright point source, the time variation of the boresight angle may be investigated, and hence the systematic offset, $\delta \alpha_{FRP_0}$, and linear drift, $\delta \alpha_{FRP_1}$, may be evaluated. 

\subsection{Evaluating the Geometric-Calibration Parameters}
\label{itr_section}

The evaluation of the reference phase, roll angle and point source positions is an iterative process, with offsets found from their inital values used in the further refinement of these parameters until convergence is reached. The offsets in the boresight angle should only need to be evaluated once, as the point source positions may be found independently of any offset in the boresight angle, as discussed above in Section~ \ref{methods_refPhase}. In practice, however, any significant offsets in the geometric-calibration parameters will affect which detections are classed as belonging to a single point source. It is therefore necessary to iterate over all the parameters, and reassign the detections as improvements in the values of the geometric-calibration parameters occur.

\section{Simulations}
\label{simul_section}

In order to assess the performance of the methods developed here in the reconstruction of the geometric-calibration parameters, it is necessary to simulate a list of point source detections for the {\it Planck} mission. This requires an input point source catalogue, together with a scanning strategy and information on the beams of the {\it Planck} detectors. Throughout this paper we have made the simplifying assumption of Gaussian beams.

\subsection{The scanning strategy}
\label{scanStrat}

While it is anticipated that this method will be applicable to any scanning strategy in which the circles corresponding to a single spin axis position may be binned together as a ring, only two potential scanning strategies for {\it Planck} were investigated here, a sinusoidal and a precessional scanning strategy. Where the sinusoidal scanning strategy may be described by:
\begin{eqnarray}
\label{Lsin_eqn}
\lambda_k & = & \lambda_0 + k \theta \nonumber \\
\beta_k & = & A \sin(n_s\lambda_k)
\end{eqnarray} and the precessional scanning strategy by:
\begin{eqnarray}
\label{Lprec_eqn}
\nu & = & (\lambda_0 + k\theta) \nonumber \\
\sin(\beta_k) & = & -\sin A \sin( n_p\nu) \nonumber \\
\cos(\phi) & = &\frac{\cos A}{\cos(\beta_k)} \nonumber \\
\lambda_k & = & \left \{ 
\begin{array}{ll} 
\nu+\phi &;\,\frac{\pi}{2} < n_p\nu < \frac{3\pi}{2}\\ 
\nu-\phi &;\,{\rm otherwise}\\ \end{array} 
\right \}
\end{eqnarray} where 
\begin{equation}
\lambda_0 = \lambda_\odot + \pi
\end{equation} where $\lambda_\odot$ is the position of the sun at the time the first ring, $k$ is the ring number, $\theta$ is the angular separation between subsequent spin axis positions, $n_{s,p}$ is the number of periods within $2\pi$, and $A$ is the amplitude. The values of these parameters used here are $\theta$=2.5\arcmin, $n_s$=2, $n_p=2.05$, and $A$=10\dg. This value of $\theta$, given a repointing once per hour, keeps the spin axis directed away from the sun.

\subsection{The input point source catalogue}
\label{inputCat}

The approximate numbers of point sources visible with {\it Planck} may be predicted by using the IRAS point source catalogue (PSC, \cite{beichman88}). This is a catalogue of some 250,000 well-confirmed point sources, providing positions, flux densities at 12, 25, 60 and 100\micron , uncertainties and various cautionary flags which are given for each source. The selection of objects from this catalogue and the extrapolation of their fluxes to Planck frequencies, is discussed in \cite{harrison04}. Also included in the input catalogue is the Wilkinson Microwave Anisotropy Probe, WMAP, point source catalogue, \cite{bennett03}. This consists of 208 extragalactic sources as seen in the WMAP maps, at {\it Planck} LFI frequencies. These sources may be extrapolated to {\it Planck} HFI frequencies using the spectral indices provided in the WMAP catalogue. The input point source catalogue constructed this way contains 5796 extragalactic point sources and 2286 galactic sources.

Table~\ref{inputCat30_table} shows the number of galactic and extragalactic sources detectable above a signal-to-noise ratio of 30 in the rings, for each of the {\it Planck} frequencies, under the assumption of the goal noise levels, shown in Table~\ref{noiseLevel_table}. 

\begin{table}
\caption{The number of galactic and extragalactic sources with fluxes greater than 30 times the nominal level of the point source sensitivity in the ring data. Numbers in italics correspond to a polarised detector pair.}
\label{inputCat30_table}
\begin{tabular}{|rrrrrrr|}
\hline
Freq & \multicolumn{6}{c}{ No. with SNR $\ge$ 30 } \\
(GHz) & \multicolumn{2}{c}{Extragalactic} & \multicolumn{2}{c}{Galactic} & \multicolumn{2}{c}{Total} \\
\hline
30 & - & {\it 19} & - &  {\it 0} & - & {\it 19} \\
44 & - & {\it 9} & - & {\it 0}  & - & {\it 9 }\\
70 & - & {\it 2} &  - & {\it 0} & - & {\it 2 }\\
100 & - & {\it 17} & - & {\it 1} & - & {\it 18} \\
143 & 24 & {\it13} & 2  & {\it 2} & 26  & {\it 15}\\
217 & 19 & {\it10} & 22  & {\it 14} & 41  & {\it 24}\\
353 & 13 & {\it5} & 99  & {\it 68} & 112 &  {\it 73}\\
545 & 21 & - & 258 &  - & 279 & -\\
857 & 81 & - & 654 &  - & 735 & - \\
\hline
{\bf Any} & \multicolumn{2}{c}{110} & \multicolumn{2}{c}{654} & \multicolumn{2}{c}{764}  \\
\hline
\end{tabular}
\end{table}

\begin{table}
\caption{The total intensity of a point source in the ring data, required for a 1 $\sigma$ detection, assuming the goal noise levels are attained.}
\label{noiseLevel_table}
\begin{tabular}{|rrr|}
\hline
Freq &  \multicolumn{2}{c}{1$\sigma$ noise (ring) (mJy)}  \\
(GHz) &  Unpolarised detector &  Polarised detector pair\\
\hline
30 & - & 147 \\
44 & - & 230 \\
70 & - & 346 \\ 
100 & - & 102 \\
143 & 85 & 120 \\
217 & 98 & 139 \\
353 & 184 & 261 \\
545 & 290 & - \\
857 & 332 & - \\
\hline
\end{tabular}
\end{table}

\subsection{Generating simulated data}
\label{simGen}

Instead of simulating time ordered data, TOD, we simulate directly the list of detections of source transits, as delivered by the Level2~DPC from their analysis of the TOD. This list of detections includes the position in phase and the amplitude observed for the transit, together with their respective errors. Additionally, the list includes the number of the detector which made the observation and the ring number on which it occurred.  

When a source will be observed by a detector depends upon when the line-of-sight of the detector passes close to the position of that source. This in turn depends upon the scanning strategy employed, the focal-plane layout and the values of the geometric calibrations parameters as discussed above. 

The simulations assume that the spin axis is repositioned every hour and that the nutation effects are small so that the individual scans may indeed be combined to reach the nominal sensitivity to point sources in rings, as shown in Table~\ref{noiseLevel_table}. Any scanning strategy which meets this proviso may be employed. The simulations also allow the variation in time of the values of the geometric-calibration parameters, as may be expected from the discussion in Section~\ref{pointingRecon}.

As any time variation in the parameters is expected to be slow, as also discussed in Section~\ref{pointingRecon}, the parameters will be constant over the time frame of an individual ring. Hence, the instantaneous offset in a geometric-calibration parameter, $\gamma$, may be defined on each ring by:
\begin{equation}
\label{inst_offset_eqn}
\gamma \left(\Gamma_i\right) = \gamma_0 + \gamma_1 \frac{\left( \Gamma_i-\Gamma_{max}/2 \right)}{\Gamma_{max}}
\end{equation} where $\Gamma_i$ is the current ring number, $\Gamma_{max}$ is the final ring of the mission and the ring numbers start from zero. The systematic offset in the parameter, $\gamma_0$, is hence defined as the instantaneous offset of the parameter exactly half-way though the mission and the drift in the value of the parameter, $\gamma_1$, is defined as the total drift in the value of the parameter over the course of the mission.

For every spin axis position, the instantaneous values of the geometric-calibration parameters are assessed and used to reconstruct the lines-of-sight of the focal plane. If a source is located nearby, the amplitude with which the source is observed is assessed and if above a threshold signal-to-noise ratio it may be included in the simulated list of detections. Errors on the amplitude of the detections are generated assuming the goal values of the noise in the ring and white noise. In order to generate realistic errors in the phase of the detection, an assessment must be made on how well the position of the transit in the scan (phase) direction may be measured. If the beams are Gaussian the position of the peak of the transit, $\psi_t$, may be found by minimising:
\begin{equation}
\label{minimise_eqn}
z(\psi_t)= \sum_i \frac{\psi_i-\psi_t}{\sigma^2_b} A_i 
\end{equation} where $\sigma_b$ is the beam sigma and $A_i$ and $\psi_i$ are the amplitudes and phases of the $i^{th}$ sample in the transit. By simulating point source transits it is possible to establish a relationship between the peak amplitude for the transit and the error in the recovered phase for the transit. Hence, an empirical relationship may be found between the signal-to-noise ratio of the detection of the point source and the magnitude of the error in the recovered phase of the transit, $\sigma_{\psi}$:
\begin{equation}
\label{phase_err_eqn}
\frac{\sigma_{\psi}}{\sigma_{b}}=\frac{1.7}{SNR} 
\end{equation} where SNR is the signal-to-noise ratio of the detection. This expression was used to determine the magnitude of the error in the phase to be included in the simulations. This phase error will dominate any errors in the phase which result from those geometric-calibration parameters determined soley by the star tracker, such as errors in the velocity-phase relation.

\section{Results}
\label{results_section}

Once a simulated list of detections has been generated, it may be analysed using the methods discussed in Section~\ref{methods_section}. The range of offsets in the geometric-calibration parameters which may be successfully recovered and the accuracies to which the recovered values may  be attained may then be investigated. Unless otherwise stated, the simulated list of detections is generated using the goal noise levels in Table~\ref{noiseLevel_table} and only includes detections from extragalactic point sources above a threshold signal-to-noise ratio of 30.

Recovering a pointing offset depends upon being able to determine which detections are due to an individual point source. The largest magnitude of pointing offset which may be recovered is therefore related to the beam sizes of the 30~GHz channel, which has the largest beams. It has proved possible to successfully recover a pointing offset of $\sim 20 \arcmin$, which is greatly in excess of the expected magnitude of the offset, as discussed in Section~\ref{pointingRecon}.

\begin{figure}
\begin{center}
\setlength{\unitlength}{1cm}
\begin{picture}(7,6.25)(0,0)
\put(8.5,-1){\includegraphics{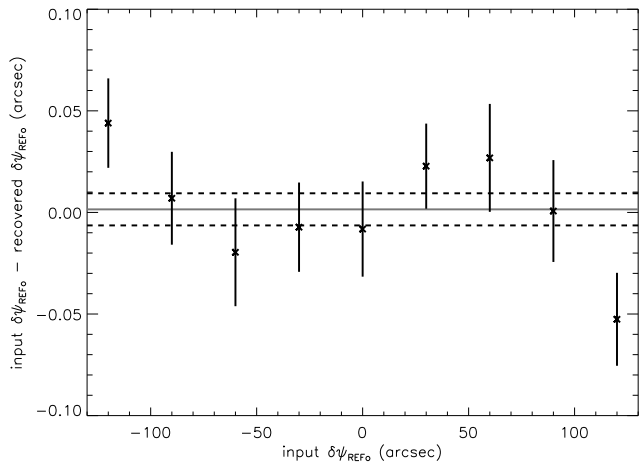}}
\end{picture}
\end{center}
\caption[]{This figure shows the ability of this method to recover systematic offsets in the reference phase, $\delta \psi_{ref_0}$, in the case of the sinusoidal scanning strategy. The differences between the input and recovered values of the systematic offset in the reference phase are plotted against the input values. For each input value, one hundred noise realisations were performed and the mean of the difference plotted with its error. The grey line is the global mean of these differences  and the dashed lines enclose the region within 1$\sigma$ of this global mean. The recovery of $\delta \psi_{ref_0}$ can hence be seen to be unbiased.}
\label{sysRefP_fig}
\end{figure}

\begin{figure}
\begin{center}
\setlength{\unitlength}{1cm}
\begin{picture}(7,6.25)(0,0)
\put(8.5,-1){\includegraphics{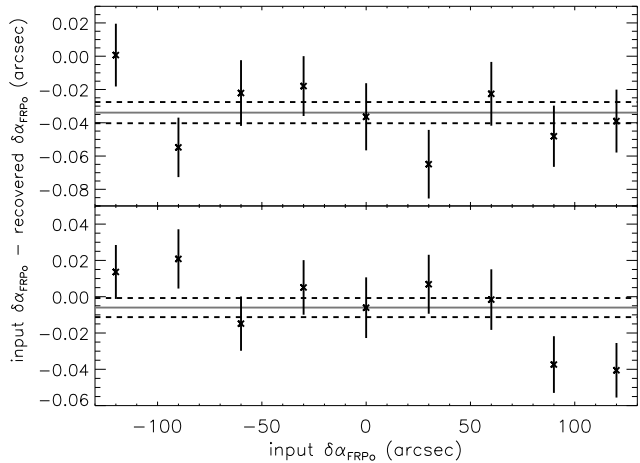}}
\end{picture}
\end{center}
\caption[]{This figure shows the difference between the input and recovered values of the systematic offset in the boresight angle, $\delta \alpha_{FRP_0}$, versus the values input to the simulations, which used the sinusoidal scanning strategy (top panel) or the precessional scanning strategy (bottom panel). For each input value, one hundred noise realisations were performed and the mean of the difference plotted with its error. The grey line is the global mean of these differences and the dashed lines enclose the region within 1$\sigma$ of this global mean. The recovery of $\delta \alpha_{FRP_0}$ is seen to be biased at the level of a few hundredths of an arcsecond in the case of the sinusoidal scanning strategy and unbiased in the case of the precessional scanning strategy.}
\label{sysBore_both_fig}
\end{figure}

Figure~\ref{sysRefP_fig} shows the difference between the value of the systematic offset in the reference phase, $\delta \psi_{ref_0}$, input to the simulations and the recovered value of $\delta \psi_{ref_0}$ against the input value, for the sinusoidal scanning strategy. This figure shows that the methods presented here have successfully recovered $\delta \psi_{ref_0}$ over a range of values which exceed those which may be expected. One hundred noise realisations per value of $\delta \psi_{ref_0}$ investigated were performed, with each mean and its associated error plotted. The grey line shows the mean value over all values of  $\delta \psi_{ref_0}$ with the dashed lines representing the 1$\sigma$ errors in this global mean. Similar analyses have been performed for all the other geometric-calibration parameters for both scanning strategies. It is found that they may be successfully recovered over the range of possible offset values. The reference phase and roll angle are found to have an unbiased recovery, but the boresight angle is particularly sensitive to the errors in the recovered point source positions and so is vulnerable to small biases in the recovery of its parameters. These biases were found to be of the order of a few hundredths of an arcsecond to a tenth of an arcsecond depending on the scanning strategy and the signal-to-noise threshold used for inclusion of detections in the analysis. Using only the highest signal-to-noise detections increases the likelihood of a biased recovery of these parameters as their assessment is then limited to a very few point sources. Given the pointing requirements, discussed in Section~\ref{acc_reqments}, a potential bias in the recovery of a geometric calibration parameter at this level is not a significant cause for concern. Figure~\ref{sysBore_both_fig} shows the recovery of the systematic offset in the boresight angle, $\delta \alpha_{FRP_0}$, for the sinusoidal and precessional scanning strategies, in the upper and lower panels respectively.  In the case of the sinusoidal scanning strategy the recovery of $\delta \alpha_{FRP_0}$ was found to be biased at the order of a few hundredths of an arcsecond, whereas no bias in the recovered value was found when the experiment was repeated using the precessional scanning strategy. 

The simulations may also be used as a check on whether the errors found for the recovered values of the geometric-calibration parameters are a true representation of the underlying errors in each of these parameters. The simulations reveal that the dispersion in the mean recovered value of a parameter is consistent with the calculated error in the parameter over the range of input values investigated, again with the exception of the boresight parameters in which the calculated error may be overestimated relative to the dispersion in the recovered value. The size of this overestimation is also found to have some dependence on the scanning strategy employed and the threshold signal-to-noise ratio used. 

\begin{figure}
\begin{center}
\setlength{\unitlength}{1cm}
\begin{picture}(7,6.25)(0,0)
\put(8.5,-1){\includegraphics{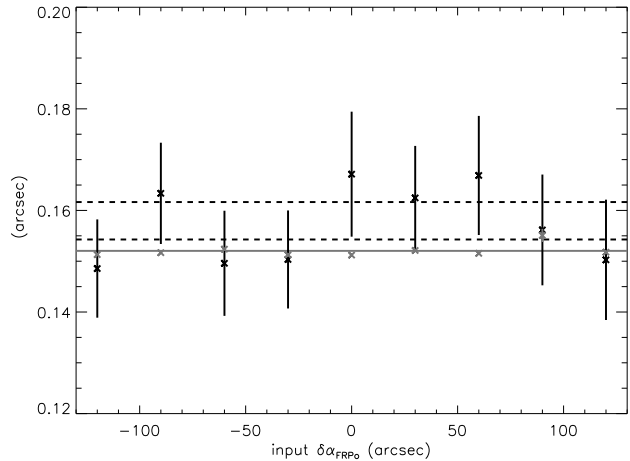}}
\end{picture}
\end{center}
\caption[]{This figure shows the dispersion in the mean recovered value of $\delta \alpha_{FRP_0}$, black crosses, and the mean calculated error in the value of $\delta \alpha_{FRP_0}$, grey crosses, against the value of $\delta \alpha_{FRP_0}$ input to the simulations, which used the precessional scanning strategy. The dashed lines enclosed the 1$\sigma$ region about the mean dispersion, and the grey line is the mean value of the calculated error. The calculated errors for $\delta \alpha_{FRP_0}$ are seen to be representative of the underlying error in the recovered value of $\delta \alpha_{FRP_0}$.}
\label{sysBoreERRs_lprec_fig}
\end{figure}

\begin{figure}
\begin{center}
\setlength{\unitlength}{1cm}
\begin{picture}(7,6.25)(0,0)
\put(8.5,-1){\includegraphics{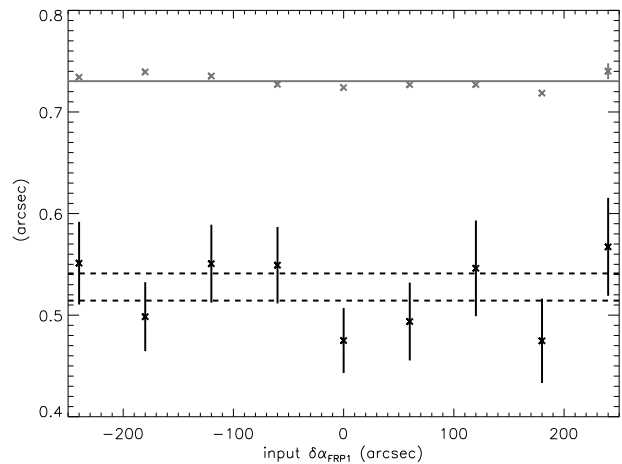}}
\end{picture}
\end{center}
\caption[]{This figure shows the dispersion in the mean recovered value of $\delta \alpha_{FRP_1}$, black crosses, and the mean calculated error in the value of $\delta \alpha_{FRP_1}$, grey crosses, against the value of $\delta \alpha_{FRP_1}$ input to the simulations, which used the precessional scanning strategy. The dashed lines enclosed the 1$\sigma$ region about the mean dispersion, and the grey line is the mean value of the calculated error. The calculated errors for $\delta \alpha_{FRP_1}$ are seen to be overestimates of the actual error in the recovered values of $\delta \alpha_{FRP_1}$, again there is no dependence of the errors on the value of $\delta \alpha_{FRP_1}$}
\label{driftBoreERRs_lprec_fig}
\end{figure}

Figure~\ref{sysBoreERRs_lprec_fig} shows the dispersion in the mean recovered value of $\delta \alpha_{FRP_0}$ and the mean calculated error in the same against the value input to the simulations, which used the precessional scanning strategy. This shows that the calculated error is representative of the actual error in the recovered value of the parameter. The error in the recovered offset also has no discernible relationship with the input value of the offset. Figure~\ref{driftBoreERRs_lprec_fig} shows the overestimation of the error in the drift of the boresight angle, $\delta \alpha_{FRP_1}$, as compared to the dispersion in the recovered value of this parameter. Again the precessional scanning strategy is used and no dependence of the errors on the initial offset in the parameter is found.

The disagreement of the errors and the occasional bias in the recovered values found for the boresight parameters are both due to errors in the recovered positions of the point sources used in this analysis. Simulations in which the correct positions of the point sources are used show no bias in the recovered values and no inconsistencies between the calculated errors and the dispersions in the recovered values. This also explains the dependence on the scanning strategy used, as this affects the errors in the recovered positions of the point sources. Limiting the evaluation of the boresight parameters to the top four frequency channels, which have the smallest beams, is found to minimise the biases in the recovered values. If required, an assessment of whether the recovered values of the boresight parameters are likely to be biased may be made using the actual point sources observed and scanning employed by {\it Planck}.

\begin{table}
\caption{Comparing the errors in the recovered values of the geometric-calibration parameters in the cases of the sinusoidal and precessional scanning strategies. In both cases only extragalactic point source detections with a signal-to-noise ratios of 30 or greater were used in the analysis.}
\label{meanCalERRS_precVsin_table}
\begin{tabular}{|lrr|}
\hline
Scanning Strategy: & sinusoidal & precessional \\
 &  (arcsec) & (arcsec) \\
\hline
$\sigma_{\psi_{ref_0}}$& 0.22 & 0.16 \\
$\sigma_{\psi_{ref_1}}$& 0.17 & 0.19 \\
$\sigma_{\alpha_{FRP_0}}$& 0.20 & 0.15 \\
$\sigma_{\alpha_{FRP_1}}$& 0.92 & 0.73 \\
$\sigma_{\rho_{0}}$&  4.2 & 3.6 \\
$\sigma_{\rho_{1}}$& 12.6 & 14.5 \\
\hline
\end{tabular}
\end{table}
Table~\ref{meanCalERRS_precVsin_table} compares the errors in the recovered values of the geometric calibrations parameters, when the sinusoidal and precessional scanning strategies are used. Due to the geometry of the ring crossings the positions of the point sources are attained to a slightly higher accuracy when the precessional scanning strategy is used, and this results in the slightly lower errors in the case of the precessional scanning strategy, especially in the case of the boresight parameters.

\begin{table}
\caption{This table shows the errors in the recovered values of the geometric-calibration parameters for different signal-to-noise ratio thresholds, for the inclusion of detections in the analysis, in the case of the sinusoidal scanning strategy.}
\label{meanCalERRS_table}
\begin{tabular}{|lrrr|}
\hline
Threshold SNR: & 40 & 30 & 20 \\
   & (arcsec) & (arcsec) & (arcsec) \\
\hline
$\sigma_{\psi_{ref_0}}$& 0.25 & 0.22 & 0.21 \\
$\sigma_{\psi_{ref_1}}$& 0.19 & 0.17 & 0.16 \\
$\sigma_{\alpha_{FRP_0}}$& 0.24 & 0.20 & 0.17 \\
$\sigma_{\alpha_{FRP_1}}$& 1.10 & 0.92 & 0.87 \\
$\sigma_{\rho_{0}}$& 4.7 & 4.2 & 3.8 \\
$\sigma_{\rho_{1}}$& 14.3 & 12.6 & 11.3 \\
\hline
\end{tabular}
\end{table}
Table~\ref{meanCalERRS_table} shows the calculated errors in the recovered values of the geometric calibrations parameters for threshold signal-to-noise ratios of 40, 30 and 20. This shows that including the lower signal-to-noise ratio detections has very little impact on the accuracies to which the geometric-calibration parameters may be attained, in the case of ideal conditions. The simulations, however, may also be used to investigate the errors in the recovered values of the geometric-calibration parameters, under non-ideal conditions. Table~\ref{meanCal30_table} shows the errors attained in the geometric-calibration parameters using all the detections with a signal-to-noise ratio above the threshold value of 30, for four different scenarios. The errors attained using the goal noise levels, shown in Table~\ref{noiseLevel_table}, are compared against those using double these goal noise levels, as well as those where the error in the phase of the peak of each transit was doubled. Given that the majority of the point sources detectable by {\it Planck}\, appear only in the top frequency channel, the above analysis was performed without this channel, to investigate whether its loss would destroy our ability to acheive the pointing requirements. The errors in the geometric-calibration parameters shown in Table~\ref{meanCal30_table}, may be expressed as errors in the pointing reconstruction. The largest errors in the reconstructed pointing will occur at the beginning and end of the mission when the errors in the drift parameters will make their largest contributions, as may be seen from equation~\ref{inst_offset_eqn}. Table~\ref{MAXpointingERR_table} shows the errors in the pointing reconstruction, as found for the errors in the geometric-calibration parameters shown in Table~\ref{meanCal30_table}, for the detectors in the HFI and LFI which have the largest uncertainty in their positions. This is due to their location in the focal plane with respect to the FRP, hence their greater sensitivity to errors in the roll angle. Table~\ref{MAXpointingERR_table} shows that even in the case of doubling the goal noise levels, the pointing requirements are still easily achievable. 
\begin{table}
\caption{The errors found in the recovered values of the geometric-calibration parameters using the sinusoidal scannning strategy and extragalactic source detections with signal-to-noise ratios greater than 30, where (i) uses the goal noise levels, (ii) excludes the 857~GHz frequency channel,(iii) doubles error in the phase of the peak of the transits and (iv) uses double the goal noise levels.}
\label{meanCal30_table}
\begin{tabular}{|lrrrr|}
\hline
  & (i) & (ii) & (iii) & (iv) \\ 
  & (arcsec) & (arcsec) & (arcsec) & (arcsec) \\
\hline
$\sigma_{\psi_{ref_0}}$& 0.22 & 0.25 & 0.45 & 0.64\\
$\sigma_{\psi_{ref_1}}$& 0.17 & 0.27 & 0.36 & 0.45 \\
$\sigma_{\alpha_{FRP_0}}$& 0.20 & 0.36 & 0.36  &0.69 \\
$\sigma_{\alpha_{FRP_1}}$& 0.92 & 1.56 & 1.71  &2.96 \\
$\sigma_{\rho_{0}}$& 4.2 & 4.3 & 8.9  &10.8 \\
$\sigma_{\rho_{1}}$& 12.6 & 12.9 & 26.2 & 36.1 \\ 
\hline
\end{tabular}
\end{table}
\begin{table}
\caption{The maximum pointing error found for an HFI and LFI detector using the errors in the recovery of the geometric-calibration parameters shown in Table~\ref{meanCal30_table}, where (i) uses the goal noise levels, (ii) excludes the 857~GHz frequency channel,(iii) doubles error in the phase of the peak of the transits and (iv) uses double the goal noise levels.}
\label{MAXpointingERR_table}
\begin{tabular}{lrrrr}
\hline
 & (i) & (ii) & (iii) & (iv) \\
 & (arcsec) & (arcsec) & (arcsec) & (arcsec) \\
\hline
HFI (545 GHz) & 0.62 & 0.95 & 1.20 & 1.93 \\
LFI (44 GHz) & 0.86 & 1.13 & 1.74 & 2.55 \\
\hline
\end{tabular}
\end{table}

\section{Discussion}

The methods presented here show that recovering the geometric-calibration parameters as part of the initial stages of the construction of the final source catalogue, successfully extracts any offsets in the values of these parameters, using only the detections due to extragalactic point sources. The accuracy to which the geometric-calibration parameters may be attained far exceeds the pointing accuracy requirements as discussed in section~\ref{acc_reqments}, so much so that it has proved possible to recover the geometric-calibration parameters under the non-ideal conditions of the loss of a frequency channel, and double the goal noise levels. The accuracies to which the geometric-calibration parameters may be recovered, are such that the errors in the pointing reconstruction due to errors in the geometric-calibration parameters are of the same level as the errors in the pointing reconstruction due to the uncertainties in the mean spin axis position recovered by the star tracker which are expected to be of the order of 1\arcsec-2\arcsec. The achievable accuracies for the geometric-calibration parameters as found above did not include errors in the remaining geometric-calibration parameters found solely from the star tracker such as errors in the mean spin axis position or the velocity-phase relation. The expected level of these errors, however, has a negligible effect on the errors found above.

The focal-plane layout, relative to the FRP, was assumed to be fixed and known. If thermal variations produce offsets in the positions of the detectors relative to the FRP, these will also need to be recovered using the science data. It is anticipated that the any positional offsets in the focal-plane layout will be recovered using the planetary transits of the focal plane. 

The methods presented here have assumed Gaussian beams. In reality, however, the {\it Planck} beams will not be Gaussian. If the beams are not symmetric about the scan direction then the phase, corresponding to the peak amplitude of the transit, will depend on the ordinate of the point source. Any dependence of the phase, of the transit, on the ordinate of the point source must be included in the analysis, and may lead to increases in the errors of the phases for the detections. We have however demonstrated that doubling the error in the phase of each transit has the effect of roughly doubling the error in the pointing reconstruction due to the geometric calibrations parameters, and these resultant errors still easily meet the required pointing accuracy. There is therefore plenty of scope for increased uncertainties in the positions of the transits.

Due to the number of available bright point sources, the time resolution of the recovery of the geometric-calibration parameters is poor and any fast evolution in a parameter will not be recoverable from the science data. Of concern is whether an unsolved variation in a parameter could bias the recovery of the parameter through an uneven distribution of detections. To investigate this possibility, the effect of an unsolved drift in the parameters was investigated. An unrecovered drift was found not to affect the recovered value of the systematic offset until the pointing requirements are exceeded by the unrecovered drift itself. It is therefore likely that any unsolved for variations will not affect the recovery of the systematic offsets and drifts while they are small enough not to be a problem in and of themselves.

\section*{Acknowledgments}
This work was supported by PPARC at the Cambridge Planck Analysis Centre.

\label{lastpage}

\end{document}